\documentclass[11pt]{article}
\usepackage{amsfonts}
\usepackage{eucal}
\usepackage{graphicx}
\usepackage{epsfig}
\begin{document}
\vskip 1cm
\noindent {\Large{\bf Elliptical Solutions to the Standard Cosmology Model with Realistic Values of Matter Density}}
\vskip 1cm

\noindent {\bf Ahmet Mecit \"Ozta\c{s}\footnote[1]{\noindent  Hacettepe Universiy Department of Physics Engineering, TR-06800, Ankara, Turkey \\ \hspace*{.4cm} e-mail: oztas@hacettepe.edu.tr},  Michael L. Smith\footnote[2]{Anabolic Laboratories, Inc. Tempe, AZ 85281, USA.  \\ \hspace*{.4cm} e-mail: mlsmith10@cox.net}}

\noindent \rule{360pt}{1pt}

{
\noindent We have examined a solution to the FRW model of the Einstein and de Sitter Universe, often termed the
 standard model of cosmology, using wide values for the normalized cosmological constant ($\Omega _{\wedge }$) and spacetime
 curvature ($\Omega _{k}$) with proposed values of normalized matter density. These solutions were evaluated using
a combination of the third type of elliptical equations and were found to display critical points for
redshift $z,$ between 1 and 3, when $\Omega _{\wedge }$ is positive. These critical points occur at values for normalized cosmological constant higher than those currently thought important, though we find this solution interesting because the $\Omega _{\wedge }$ term may increase in dominance as the Universe evolves bringing this discontinuity into
importance.  We also find positive $\Omega _{\wedge }$ tends towards attractive at values of $z$ which are commonly
observed for distant galaxies.}

\noindent \rule{360pt}{1pt}

\vskip .5cm 

\noindent {Key words~~~~: }{\it cosmology, cosmological constant, FRW,  vacuum energy,

\hskip 2cm redshift}


\section{INTRODUCTION}

The standard view of Universe expansion, the Friedmann-Robertson-Walker (FRW) model, seems to
 mimic our current situation well given the approximation of an isotropic and homogeneous sprinkling of
 matter across large dimensions and inclusion of vacuum energy. The standard model has several variants
 and most include significant positive values of $\wedge$ the vacuum energy constant; this energy of deep space
 seems to encourage Universe expansion and act as anti-gravity. Some current detailed models predict a
 greatly increasing importance of this energy with continued Universe expansion, with suggestions we are
 entering the age of dominance by vacuum energy (Behar and Carmeli, 2000). Another recent, interesting
 model details such an energy coupled to absolute time, allowing an initial vacuum energy driven
 inflationary phase immediately after release from singularity, followed by a more leisurely relaxation of
 this energy (Bisabr, 2004 and references therein). Perhaps the current vacuum energy is a residual of the
 first moments of the Universe.
\vskip .3cm
\noindent Recent measurements of supernovae type Ia (SNe Ia) distances D$_{L}$ and associated redshifts $z$ (or
receding velocities), have uncovered the existence of a significant current value for the cosmological constant
(Tonry {\em et al.,} 2003, Reiss {\em et al., }2004, 1998). The measurements of SNe Ia distances and receding 
velocities are subject to a myriad of possible experimental errors demanding careful data collection and 
detailed analyses, for evidence of this positive repulsive energy is obtained from quite distant SNe Ia 
explosions. Still these SNe Ia studies present data with the smallest experimental errors to date and are 
the most useful for model testing. These authors also take the positive $\wedge$ to mean the Universe is 
expanding more rapidly now than in the past and the local Hubble constant, $H_{0}$ is gradually increasing 
with time. Without this energy from deep space we should expect $H_{0}$ to slowly decrease over time due to 
the self attraction of matter and energy. Knowledge of $H_{0}$ is also important because the upper bound of 
the age of the Universe is fixed by the local value of $H_{0}$ in many models. Surprisingly, $H_{0}$ is not known 
to better than $\pm $10\% (Freedman, {\em et al.,} 2001), indeed the Universe age is not known to better than 
$\pm $10\% if the age of low metal stars is a gauge (VandenBerg, {\em et al.,} 2002, Grundahl, {\em et al.,} 2000). One 
reason for this is the dependence of $H_{0}$ upon matter density which itself is highly dependent upon the 
distances between stars and galaxies and dust density. Another reason for inaccuracy being the relative 
time span of our data, both SNe Ia and star elemental composition, are instantaneous respective to 
processes of the Universe.
\vskip .3cm
\noindent We have examined the standard model of the expanding Universe using a solution of the elliptical form
 and found that at realistic matter densities this solution exhibits discontinuities at values of the
 normalized cosmological constant slightly larger than current estimates. When we examine this solution for the
 smaller normalized matter densities of the future (and values suggested a few decades ago), this
 discontinuity presents problems for vacuum energy dominated models. Use of the cosmological constant in the
 standard model also clouds predictions of galactic redshifts at epochs only slightly older than those
 collected in SNe Ia studies. Though it would be useful to be able to calculate distant galactic ages from
 redshift alone, now at 7 or higher (Kneib, {\em et al.,} 2004), such are unreliable if the positive vacuum energy
 is significant. We do not find discontinuities for negative vacuum energies (attractive) at realistic values
 of normalized matter densities.

\section{THE FRIEDMANN MODEL}

\noindent We use the conventions of Carroll {\em et al.} (1992) with a Friedmann equation modeling our expanding
 Universe
\begin{equation}H^{2} = (\frac{\dot {R}}{R})^{2} = \frac{8\pi G}{3}\rho _{m} + \frac{\wedge }{3} - \frac{k}{R^{2}}.\end{equation}
Here the $\rho _{m}$ is the matter density and the first term attractive, while the second term may either be
 repulsive or attractive, it is usually meant as repulsive. The constant of integration $k,$ may take values of
 -1, 0 or +1, for a Universe with "open, flat or closed" geometries, respectively. The $H$ represents the
 Hubbell parameter with $R$ the expansion factor for the evolving Universe.
\vskip .3cm
\noindent Einstein first proposed his gravitational equation without a cosmological constant and preferred a very slightly
closed Universe with matter dominating (Einstein, 1915), that is $k >$ 0. He later introduced the cosmological
constant into his gravitational equation after Friedmann and Lema$\hat {\mbox{i}}$tre pointed out the Universe,
populated with considerable matter, should either be expanding or suffer contraction, but astronomers
had not yet firmly discovered other "Universes" or galaxies outside the Milky Way. This constant
allowed the possibility of a static Milky Way (and Universe), which was the limit of knowledge early in
the 20th century. Hubble, Wirtz, Slipher and others later pointed out that most other galaxies were
following trajectories away from the Milky Way with the implication that the Universe has no possible
 stationary reference point, in confirmation of Einstein's proposals. It seems that Einstein later regretted
 introduction of this cosmological constant, nonetheless, this concept has recently regained popularity in
 cosmology to explain certain observations.

\section{THEORY}

It is common to introduce normalized parameters for matter, vacuum energy and geometry
\begin{equation}\Omega _{m} = \frac{8\pi G}{3H_{o}^{2}}\rho _{m_{o}} , \hskip0.2cm    \Omega _{\wedge } = \frac{\wedge }{3H_{o}^{2}} ,\hskip0.2cm    \Omega _{k} = -\frac{k}{R_{o}^{2}H_{o}^{2}}.\end{equation}

\noindent We use the typical conditions of normalization across these three parameters following the convention of
 Carroll {\em et al.} (1992)
\begin{equation}1 = \Omega _{m} + \Omega _{\wedge } + \Omega _{k}\end{equation}
\noindent where $\Omega _{m}$ represents normalized matter density, $\Omega _{\wedge }$ normalized vacuum energy density and $\Omega _{k}$ 
normalized (possible) spacetime curvature; the radiation density term, $\Omega _{\mbox{r}}$ being small at present, is
 included with matter density and we will briefly review the equations of interest. 
If we allow $\rho _{m_{o}}$ to represent the present matter density which is $\frac{M}{(4\pi /3R_{o}^{3})}$ in equation (1) to give us the FRW model at the present time we have
\begin{equation}(\frac{dR}{Rdt})^{2} = H_{o}^{2}\Omega _{m}\frac{R_{o}^{3}}{R^{3}} + H_{o}^{2}\Omega _{\wedge }  + H_{o}^{2}\Omega _{k}\frac{R_{o}^{2}}{R^{2}} .\end{equation}

\noindent Now we substitute for $\frac{R}{R_{o}}$ with $a(t)$
\begin{equation} (\frac{\dot {a}(t)}{a(t)})^{2} = H_{o}^{2}\frac{\Omega _{m}}{a(t)^{3}} + H_{o}^{2}\Omega _{\wedge }  + H_{o}^{2}\frac{\Omega _{k}}{a(t)^{2}}\end{equation}
\noindent and then multiplying through by $\frac{a(t)^{2}}{H_{o}^{2}}$ gives us one equation of our current state
\begin{equation} \frac{1}{H_{o}^{2}}(\frac{da(t)}{dt})^{2} = \frac{\Omega _{m}}{a(t)} + a(t)^{2}\Omega _{\wedge } + \Omega _{k} .\end{equation}
\noindent This allows us to introduce the dimensionless parameter $\tau $ for time as $\tau  = H_{o}t$
\begin{equation}(\frac{da(\tau )}{d\tau })^{2} =  \frac{1}{a(\tau )}(\Omega _{m} + \Omega _{\wedge }a(\tau )^{3} + \Omega _{k}a(\tau ))\end{equation}
\begin{equation}\frac{da(\tau )}{d\tau } =  \frac{1}{\sqrt{a(\tau )}} \sqrt{\Omega _{m}+ \Omega _{\wedge }a(\tau )^{3} + \Omega _{k}a(\tau )}.\end{equation}
\noindent Then inverting and integrating both sides from the past $\tau _{1}$ to the present $\tau _{o}$ we have
\begin{equation}\int_{\tau _{1}}^{\tau _{0}} 1 d \tau =  \int_{a_{1}}^{1} \frac{\sqrt{a(\tau )}}{\sqrt{\Omega _{m}+ \Omega _{\wedge }a(\tau )^{3} + \Omega _{k}a(\tau )}} d a(\tau )\end{equation}

\noindent and substituting $1/(1+z)$ for $a$ we have the integral from the past $z_{1}$ to the present of 0 we arrive at
\begin{equation}\tau _{o} - \tau _{1} = - \int_{z_{1}}^{0} \frac{1}{(1+z)\sqrt{\Omega _{m}(1+z)^{3}+\Omega _{\wedge }+\Omega _{k}(1+z)^{2}}} d z .\end{equation}

\noindent This is similar to an equation modeling redshifts as presented in Peebles (1993). We shall use the variable
 $y$ for $1+z$ and bring out $\frac{1}{\sqrt{\Omega _{m}}}$ from within the integral of (10), so the integral
 becomes 
\begin{equation}\tau _{o} - \tau _{1} = - \frac{1}{\sqrt{\Omega _{m}}}\int_{1+z_{1}}^{1} \frac{1}{y\sqrt{y^{3} +\frac{\Omega _{\wedge }}{\Omega _{m}} +\frac{\Omega _{k}y^{2}}{\Omega _{m}}} } d y .\end{equation}

\noindent We shall now change the variable once again, allowing $y = u - \frac{\Omega _{k}}{3}\Omega _{m} $which also changes the  integration limits of the following equation
\begin{equation}\tau _{o} - \tau _{1} = \frac{1}{\sqrt{\Omega _{m}}}\int_{1+\frac{\Omega _{k}}{3\Omega _{m}}}^{1+z_{1}+\frac{\Omega _{k}}{3\Omega _{m}}} \frac{1}{(u - \frac{\Omega _{k}}{3\Omega _{m}})\sqrt{u^{3} - \frac{\Omega _{k}^{2}}{3\Omega _{m}^{2}}+\frac{2\Omega _{k}^{3}}{27\Omega _{m}^{3}}+\frac{\Omega _{\wedge }}{\Omega _{m}}}} d u\end{equation}

\noindent and by using a similar substitution we can simplify the denominator in terms of $u$ with the same limits of
 integration as equation (12)
\begin{equation}\tau _{o} - \tau _{1} = \frac{1}{\sqrt{\Omega _{m}}}\int \frac{1}{(u - \frac{\Omega _{k}}{3\Omega _{m}})\sqrt{(u-u_{1})(u-u_{2})(u-u_{3})}} d u .\end{equation}

\noindent The right side of equation (13) is solved as an elliptical function of the first and third type, where
\begin{equation}x = \sqrt{\frac{u-u_{1}}{u_{2}-u_{1}}}, \nu  = \frac{u_{1}-u_{2}}{(u_{1} - \frac{\Omega _{k}}{3\Omega _{m}})}, k = \sqrt{\frac{u_{1}-u_{2}}{u_{1}-u_{3}}}\end{equation}
\begin{equation}\tau _{o} - \tau _{1} = - \frac{2}{\sqrt{u_{3}-u_{1} }(u_{3} -\frac{\Omega _{k}}{3\Omega _{m}})}\times\Pi (x,\nu ,k)\end{equation}

\noindent and the function can be evaluated using the limits of Eq. (12). We may also proceed from equation (10) by inverting the limits of integration, changing sign and substituting for $\tau _{o} - \tau _{1}$ we get the following useful equation with details presented in the appendix
\begin{equation}H_{0}D_{L} = \frac{c(1+z)}{\sqrt{\left| \Omega _{k}\right| }} \mbox{sin}n\{\frac{\sqrt{\left| \Omega _{k}\right| }}{\sqrt{\Omega _{m}}}\int_{0}^{z_{1}} \frac{1}{\sqrt{(1+z)^{3}+\frac{\Omega _{\wedge }}{\Omega _{m}}+\frac{\Omega _{k}}{\Omega _{m}}(1+z)^{2}}} d z\end{equation}

\noindent and sin$n$ is sin for $\Omega_{k} < 0$ and sinh for $\Omega_{k} >0$. Note that within the denominator of the integral $\Omega_{\wedge}$ appears in only linear combination with z as opposed to appearance of this term in the more conventional use (Carroll {\em et al.} 1992, Tonry {\em et al.,} 2003) in this equation
\begin{equation}H_{0}D_{L} = \frac{c(1+z)}{\sqrt{\left| \Omega _{k}\right| }}\mbox{sin}n\{\left| \Omega _{k}\right| ^{1/2}\int_{0}^{z_{1}} \frac{1}{\sqrt{(1+z)^{2}(1+\Omega _{m}z)-z(2+z)\Omega _{\wedge }}} d z.\end{equation}

\noindent Again, the nature of sin$n$ depends upon spacetime curvature as above and equations (16) and (17) are
 different forms of the same solution.

\section{RESULTS}

\noindent There are some problems of inconsistency that arise with this solution of the standard model and we
 observe these near parameter values presently considered important. We find the root portion of the
 denominator of (10) causes these inconsistencies across a range of normalized matter densities as
 displayed in Figure 1 for $\Omega _{m}$ from $0.30$ to 0.05. These graphs present the discontinuity, with respect to
 equation (10), increasing in intensity with corresponding value of $z$ as $\Omega _{m}$ approaches 0. For instance, at
$\Omega _{m}$ of 0.30, currently of interest (Feldman {\em et al.,} 2003), $\Omega _{\wedge }$ appears inconsistent from 1.713 to 1.714 with the associated values of $\Omega _{k}$ of -1.013 and -1.014. This inconsistency widens considerably at $\Omega _{m}$ of 0.05, perhaps of interest (Fall, 1975), where at z $\approx $ 2.65 the inconsistency in $\Omega _{\wedge }$ lies from 1.224 to 1.250; though the breadth of the inconsistency increases with increasing $z$ and decreasing $\Omega _{\wedge }$ the exact values of these inconsistencies decrease along with disappearing matter density.

\begin{figure}[ht]
\hspace*{-1cm}
\includegraphics[width=1.1\textwidth]{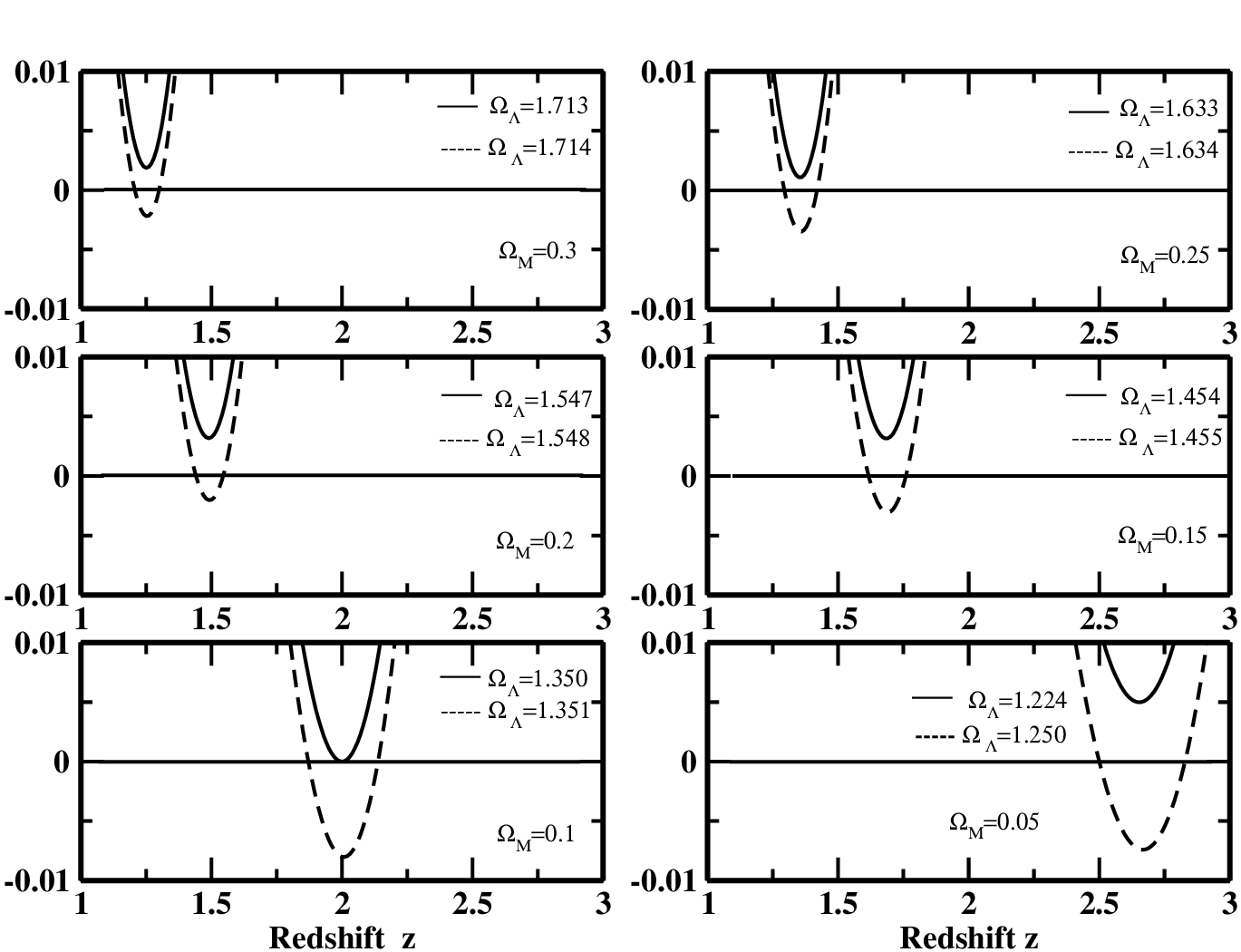}
\caption{\{$\Omega _{m} (1+z)^{3} + \Omega _{\wedge } + \Omega _{k} (1+z)^{2}\}$ versus redshift $z$ for various normalized matter densities between 0.30 and 0.05.}
\end{figure}
 
\begin{figure}[ht]
\hspace*{-1cm}
\includegraphics[width=1.1\textwidth]{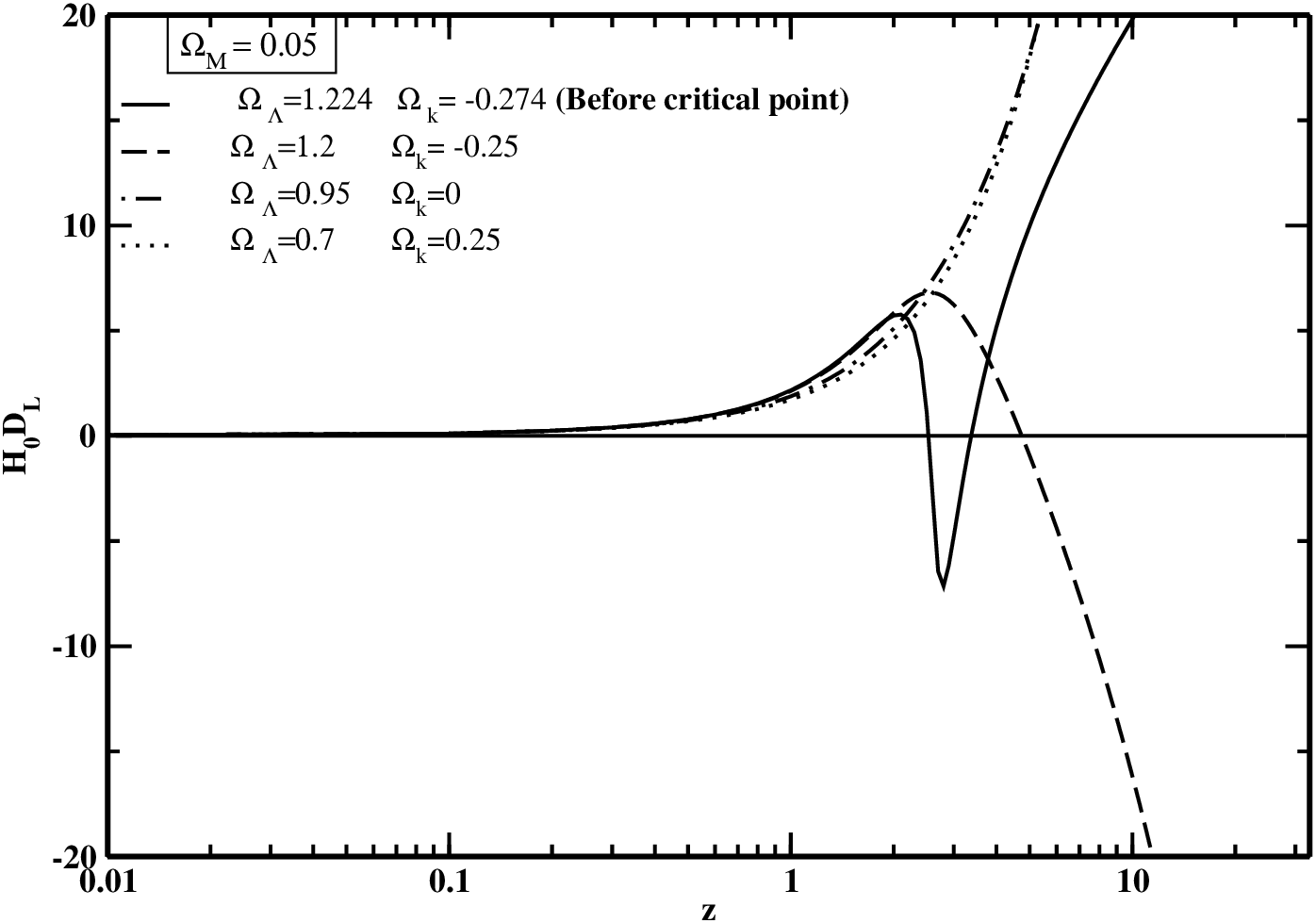}
\caption{Plots of $H_{0}D_{L}$ as functions of log $z$ with various values of $\Omega _{\wedge }$ with $\Omega _{m}$ of 0.05 and c normalized to 1.}
\end{figure}

\noindent We have traced these inconsistencies at $\Omega _{m}$ of 0.05 and 0.30, with various values for $\Omega _{\wedge }$ and present results in Figures 2 and 3 as plots of $H_{0}D_{L}$ with respect to log$z$. Both figures demonstrate for large values of the cosmological constant, 1.2 to 1.7, and with reasonable values of spacetime curvature $\Omega _{k}$ drastic deviations from useful solutions of $H_{0}D_{L}$. At the lower matter density of 0.05 (Fig. 2) this function does
not straightforwardly correlate $H_{0}D_{L}$ with $z$, for values of $\Omega _{\wedge }$ above 1.2 and for $z$ above about 2. Here
the critical value for $\Omega _{\wedge }$ is a rather low 1.224. With increasing $\Omega _{\wedge }$ from 0.70 to 0.95, we find the larger
values of $z$ correlate with the smaller values of $\Omega _{\wedge }$ as should be observed in a Universes with increasing
matter densities and decreasing anti-matter densities.
\vskip .3cm
\noindent
As plotted in Fig. 3 ($\Omega _{m}$= 0.30), all curves are shifted to higher values of z, as expected for a Universe
more populated with matter than in Figure 2. The model appears well behaved with larger values of
$H_{o}D_{L}$ correlating with larger values of redshift up to perhaps $\Omega _{\wedge}$ of 1.2; solutions of $H_{o}D_{L}$ at values of $\Omega _{\wedge}$ greater than this and above $z$ of 3.25 are questionable. Here, evaluations of $H_{0}D_{L}$ with smaller values of $\Omega _{\wedge }$ appear indiscriminate but perhaps useful up to $z$ of about 3, but no further. Beyond this redshift $z$ is predicted to correlate with anti-gravity density. Though these problems of inconsistency are for $\Omega _{\wedge }$ greater than 0.70 - just outside the range currently thought to be important - as the Universe expands, $\Omega _{\wedge }$ is supposed to dominate with $\Omega _{m}$ eventually approaching 0 and the standard model may fail for even moderate values of the redshift and $\Omega _{\wedge}$.
\vskip .3cm
\noindent We have found the FRW model does not obviously fail for negative values of $\Omega _{\wedge}$ over a much wider
range of $\Omega _{m}$ and $z$. Figure 4 presents log $H_{0}D_{\mbox{L}}$ versus log$z$ for $\Omega _{m}$ of 0.05 with $\Omega _{\wedge}$ from +1 to -2. As expected for a Universe dominated by attractive forces, $H_{0}D_{L}$ decreases with decreasing $\Omega _{\wedge}$ and $z$ decreases with decreasing $\Omega _{\wedge}$ in smooth manners because the Hubble flow must increase with increasing drag on expansion. Note that a positive cosmological constant ($\Omega _{\wedge}$ = 1) does follow this regular pattern above redshift of 6.59 in the Universe of low matter density (0.05); greater than this distance the energy is predicted to become strongly attractive. This ordering also exists at $\Omega _{m}$ of 0.30 (Fig. 5) but is more pronounced; up to redshifts of about 3.25 a $\Omega _{\wedge}$ of +1.0  acts as anti-gravity, but at greater distance than this $\Omega _{\wedge}$ acts attractively. 

\begin{figure}[ht]
\hspace*{-1cm}
\includegraphics[width=1.1\textwidth]{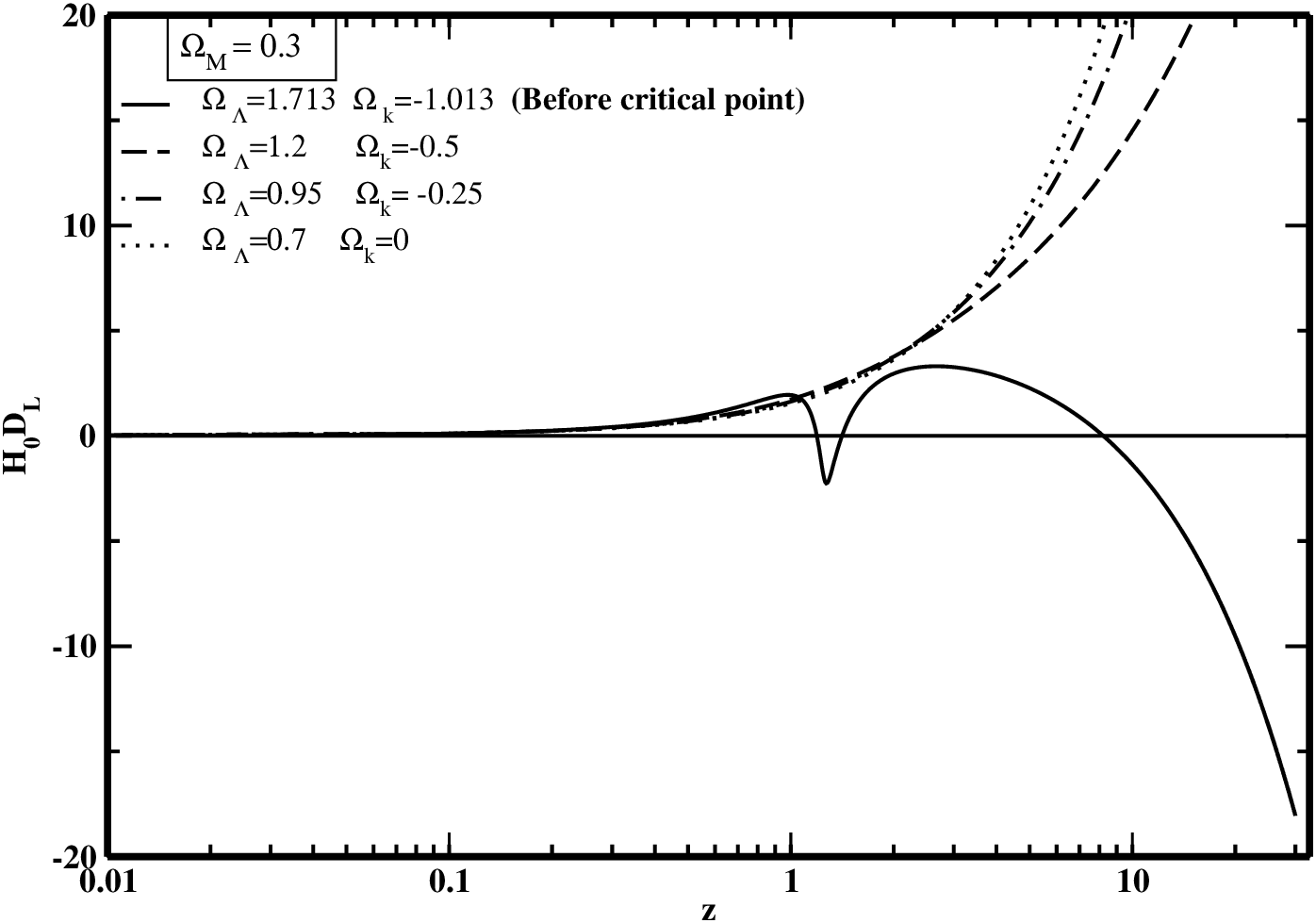}
\caption{Plots of $H_{0}D_{L}$ as functions of log $z$ with various values of $\Omega _{\wedge }$ with $\Omega _{m} $of 0.30 and c normalized to 1.}
\end{figure}

\begin{figure}[ht]
\hspace*{-1cm}
\includegraphics[width=1.1\textwidth]{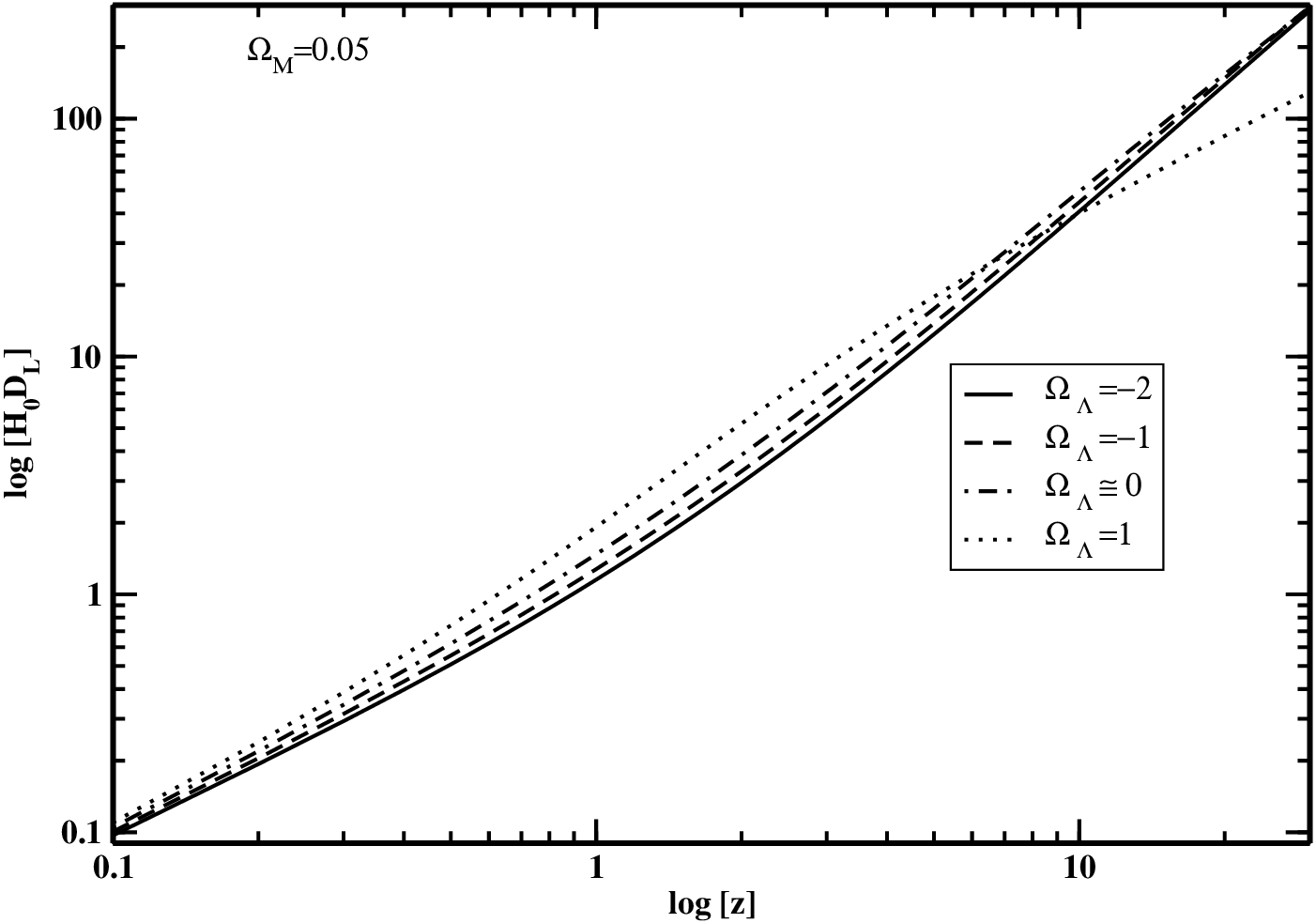}
\caption{Log $H_{0}D_{L}$ as a function of log z for various values of $\Omega _{\wedge }$ from +1.0 to -2.0 with $\Omega _{m}$ of 0.05 and c normalized to 1. Note the "cross-over" of the trace of the solution with $\Omega _{\wedge }$ of +1.0.}
\end{figure}

\begin{figure}[ht]
\hspace*{-1cm}
\includegraphics[width=1.1\textwidth]{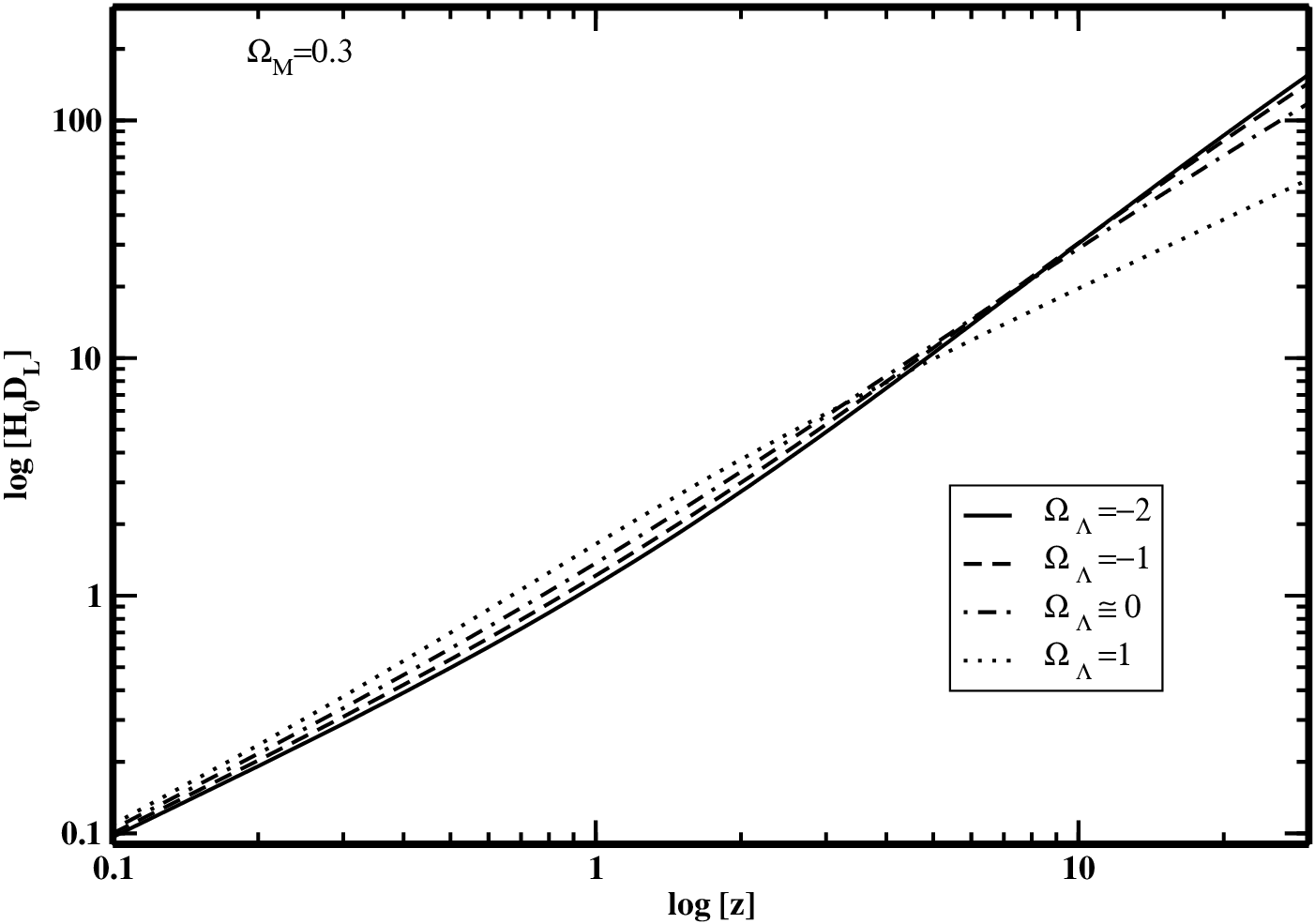}
\caption{Log $H_{0}D_{L}$ as a function of log z for various values of $\Omega _{\wedge }$ from +1.0 to -2.0 with $\Omega _{m}$ of 0.30 and c normalized to 1.  Note the "cross-over" of the trace of the solution with $\Omega _{\wedge }$ of +1.0.}
\end{figure}

\section{DISCUSSION}

The selection of a solution to the standard model of cosmology using an elliptical form allows us to probe
the ability of the model to estimate Hubble expansion at great distances with dependence upon $\Omega _{\wedge}$. We
have found, for values of $\Omega _{\wedge}$ between 0 and -2 and values of $\Omega _{m}$ currently thought realistic, the
standard model returns a consistent value for $H_{0}D_{L}$ at $z <$ 20. This is greater than redshifts currently
reported for the farthest galaxies, though this value may be surpassed soon. When this model is solved
for values of $\Omega _{\wedge}$ greater than 0 problems arise at high values of $z$, where positive $\Omega _{\wedge}$ acts as attractive. In regions of $\Omega _{\wedge}$ greater than +1 inconsistencies are observed of a general nature which limits the usefulness of the standard model to regions of high matter density - densities consistent with the
Einstein-de Sitter model and currently thought important. If the Universe is rushing towards a state of
low matter density, increasing dependence upon $\Omega _{\wedge}$ with a trace of closed curvature (preferred by
Einstein among others) the standard model will probably fail. These regions of inconsistencies might be
interpreted by some, as evidence for an epoch of matter dominated Universe with a low Universe
expansion rate followed by the current epoch of vacuum energy domination and faster expansion,
though it seems to us that earlier epochs might be best not judged using the standard model. Very
unfortunately, this also means calculations of epochs of newly discovered galaxies exhibiting high
redshifts will remain unreliable until the FRW model can be better adapted for greater distances.
The standard model has limitations of usefulness with respect to accurate predictions of large galactic
distances and Universe age. Inclusion of positive values for the vacuum energy further
restricts the range of useful predictions. Other models or model variations incorporating modified
application of the vacuum energy, but not suffering regions of inconsistency, might be preferred. We
hope to introduce one such model soon.

\vskip1.5cm

\noindent {\bf ACKNOWLEDGEMENTS}
\vskip .3cm
\noindent We are grateful for the continuing interest of Professor Jan Paul in our work.

\vskip .3cm

\noindent {\bf REFERENCES}
\vskip .3cm
\noindent Behar, S. and Carmeli, M. (2000) {\em International Journal of Theoretical Physics} {\bf 39,} 1375.
\vskip .1cm
\noindent Bisabr, Y. (2004) {\em International Journal of Theoretical Physics} {\bf 43,} 2137.
\vskip .3cm
\noindent Carroll, S.M., Press, W.H. and Turner, E.L. (1992) {\em Annual Review of Astronomy and Astrophysics} {\bf 30,} 499.
\vskip .3cm
\noindent Einstein, A. (1916) {\em Relativity,} translation 2000, Routledge Classics, NY.
\vskip .3cm
\noindent Fall S.M. (1975) {\em Monthly Notices of  the Royal Astronomical Society} {\bf 172, }23.
\vskip .3cm
\noindent Feldman, H. Jusziewicz, R., Ferreira, P., Davis, M., Gaztanas, E., Fry, J., Jaffee, A., Chambers, S., La Costa, L., Bernardi, M., Giovanelli, R., Haynes, M. and Wegner, G. (2003) {\em Astrophysical Journal} {\bf 596,} L131.
\vskip .3cm
\noindent Freedman, W.L., Madore, B.F., Gibson, B.K., Ferrarese, L., Kelson, D.D., Sakai, S., Mould, J.R., Kennicutt, Jr., R.C., Ford, H.C., Graham, J.A., Huchra, J.P., Hughes, S.M.G., Illingworth, G.D., Macri, L.M. and Stetson, P.B. (2001) {\em Astrophysical Journal} {\bf 553,} 47.
\vskip .3cm
\noindent Grundahl, F. VandenBerg, D.A., Bell, R.A., Andersen, M.I. and Stetson, P.B. (2000) {\em Astronomical Journal} {\bf 120,} 1884.
\vskip .3cm
\noindent Kneib, J.-P., Ellis, R.S., Santos, M.R. and Ricard J. (2004) {\em Astrophysical Journal }{\bf 607, }697.
\vskip .3cm
\noindent Peebles, P.J.E. (1993) {\em Principles of Physical Cosmology, }Princeton University Press, Princeton, New Jersey, pp.100-102.
\vskip .3cm
\noindent Riess, A.G., Strolger, L-G., Tonry, J., Casertano, S., Ferguson, H.C., Mobasher, B., Challis, P., Filippenko, A.V., Jha, S., Li, W., Chornock, R., Kirshner, R.P., Leibundgut, B., Dickinson, M., Livio, M., Giavalisco, M., Steidel, C.C., Benitez, N., and Tsvetanov, Z. (2004). {\em Astrophysical Journal} {\bf 607,} 665.
\vskip .3cm
\noindent Riess, A.D., Filippenko, A.V., Challis, P., Clocgiatti, A., Dierks, A., Garnavich, P.M., Gilliland, R.L., Hogan, C.J., Jha, S., Kirshner, R.P., Leibundgut, B., Phillips, M.M., Reiss, D., Schmidt, B.P., Schommer, R.A., Smith, R.C., Spyromilio, J., Stubbs, C., Suntzeff, N.B. and Tonry, J. (1998) {\em Astronomical Journal} {\bf 116,} 1009.
\vskip .3cm
\noindent Tonry, J.L., Schmidt, B.P., Barris, B., Candia, P., Challis, P., Clocchiatti, A., Coil, A.L., Filippenko, A.V., Garnavich, P., Hogan, C., Holland, S.T., Jha, S., Kirshner, R.P., Krisciunas, K., Leibundgut, B., Li, W., Matheson, T., Phillips, M.M., Riess, A.G., Schommer, R., Smith, R.C., Sollerman, J., Spyromilio, J., Stubbs, C.W. and Suntzeff, N.B. (2003) {\em Astrophysical Journal} {\bf 594,} 1.
\vskip .3cm
\noindent VandenBerg, D.O., Ricard, O., Michaud, G. and Richer J. (2002) {\em Astrophysical Journal} {\bf 571,} 487.

\subsection*{\bf APPENDIX}

\noindent We derive equation (16), the elliptical form useful for astronomy from our equation (9). The FRW  metric, allowing $a(t) = \frac{R}{R_{o}}$ is
\begin{equation}\frac{dR_{o}r}{dt} = \frac{R_{o}}{R}(1-kr^{2})^{1/2}\end{equation}
\begin{equation}R_{o}\frac{dr}{dt} = \frac{1}{a(t)}(1-kr^{2})^{1/2}\end{equation}

\noindent with rearrangement becomes
\begin{equation}\frac{R_{o}dr}{(1-kr^{2})^{1/2}} = \frac{dt}{a(t)}.\end{equation}

\noindent Remembering that $\tau  $= $H_{o}t$ we multiply through by $H_{o} $to get
\begin{equation}H_{o}R_{o}\frac{dr}{(1-kr^{2})^{1/2}} = \frac{d\tau }{a(t)}\end{equation}

\noindent and multiply through again using the relationship for the normalized $\Omega _{k}=-\frac{k}{R_{0}^{2}H_{o}^{2}} $
\begin{equation}H_{o}R_{o}\frac{dr}{(1+\Omega _{o}R_{o}^{2}H_{o}^{2}r^{2})^{1/2}} = \frac{d\tau }{a(t)}.\end{equation}

\noindent Substituting equation (9) for $d\tau $ we have 
\begin{equation}H_{o}R_{o}\frac{dr}{(1+\Omega _{o}R_{o}^{2}H_{o}^{2}r^{2})^{1/2}} = \frac{1}{a(\tau )}\frac{  \sqrt{a(\tau )} da(\tau )}{\sqrt{\Omega _{m} + \Omega _{\wedge }a(\tau )^{3} + \Omega _{k}a(\tau )}}\end{equation}

\noindent and integrating both sides and rearranging some constant parameters
\begin{equation}H_{o}R_{o}\int_{0}^{r_{1}} \frac{1}{\sqrt{1+\Omega _{o}R_{o}^{2}H_{o}^{2}r^{2}}} d r = \int_{a_{1}}^{1} \frac{1}{\sqrt{a(\tau )}\sqrt{\Omega _{m} + \Omega _{\wedge }a(\tau )^{3} + \Omega _{k}a(\tau )}} d a(\tau )\end{equation}

\noindent then changing the variable on the left side using $y = \sqrt{\Omega _{k}}R_{o}H_{o}r$ and using the redshift relationship $a =$  $\frac{1}{(1+z)}$ for substitution of the right hand side we get 
\begin{equation}\frac{1}{\sqrt{\Omega _{k}}} \int_{0}^{\sqrt{\Omega _{k}}R_{0}H_{0}r_{1}} \frac{1}{\sqrt{1+y^{2}}} d y = - \int_{z_{1}}^{0} \frac{\sqrt{1+z}}{(1+z)^{2}\sqrt{\Omega _{m} + \frac{\Omega _{\wedge }}{(1+z)^{3}}  +  \frac{\Omega _{k}}{1+z} }} d z.\end{equation}

\noindent The integral on the left hand side is $ \textrm{\rm arcsinh}(y)$ so the above becomes
\begin{equation}\frac{1}{\sqrt{\Omega _{k}}}\textrm{\rm arcsinh} (\sqrt{\Omega _{k}}R_{0}H_{0}r_{1}) = \int_{0}^{z_{1}} \frac{1}{\sqrt{\Omega _{m}(1+z)^{3} + \Omega _{\wedge } + \Omega _{k}(1+z)^{2}}} d z\end{equation}

\noindent and substituting to arrive at the measurable $D_{L}$ using the relationships $D_{m} = R_{o}r_{1}$ and $D_{L} = (1+z)D_{m} $
\noindent we arrive at 
\begin{equation}H_{o}D_{L} = \frac{1+z}{\sqrt{\left| \Omega _{k}\right| }}\sin \mbox{\bf n} \{\frac{\sqrt{\left| \Omega _{k}\right| }}{\sqrt{\Omega _{m}}}\int_{0}^{z_{1}} \frac{1}{\sqrt{(1+z)^{3}+ \frac{\Omega _{\wedge }}{\Omega _{m}}+ \frac{\Omega _{k}}{\Omega _{m}}(1+z)^{2}} } d z\end{equation}

\noindent and with introduction of c for the speed of light becomes our equation (16).

\end{document}